\documentclass[sigconf]{acmart}
\usepackage{url}
\usepackage{hyperref}
\usepackage{booktabs} % For formal tables
\usepackage{amsmath}
\usepackage{amsfonts}
\usepackage{comment}
\usepackage{soul}
\usepackage{mathrsfs}
\usepackage{graphicx}
\usepackage{subfigure}
\usepackage{multirow}
\usepackage{algorithm}
\usepackage{algorithmic}
\usepackage{multicol}  
\usepackage{enumitem}
\usepackage{array}
\usepackage{float}
\usepackage{hyperref} % For clickable email links
\usepackage{tcolorbox}  % 文本加框
\tcbuselibrary{breakable}   % 配合文本加框，使得框可以跨页

\AtBeginDocument{%
	\providecommand\BibTeX{{%
			\normalfont B\kern-0.5em{\scshape i\kern-0.25em b}\kern-0.8em\TeX}}}
\copyrightyear{2025}
\acmYear{2025}
\setcopyright{acmlicensed}
\acmConference[CIKM '25]{Proceedings of the 34th ACM International Conference on Information and Knowledge Management}{ November 10--14, 2025}{Seoul, Republic of Korea.} % {October 14–-18, 2024}{Bari, Italy}
\acmBooktitle{Proceedings of the 34th ACM International Conference on Information and Knowledge Management (CIKM '25), November 10--14, 2025, Seoul, Republic of Korea}
\acmISBN{979-8-4007-2040-6/2025/11}
\acmDOI{10.1145/3746252.3761574}
\begin{document}

\title{Prompt Tuning as User Inherent Profile Inference Machine}

\author{Yusheng Lu}
\authornote{Yusheng Lu and Zhaocheng Du are co-first authors with equal contribution.}
\affiliation{
\institution{Tongji University}
%\institution{City University of Hong Kong}
\state{Shanghai}
\country{China}
}
\affiliation{
%\institution{Tongji University}
\institution{City University of Hong Kong}
\state{Hong Kong}
%\state{Shenzhen}
\country{China}
}
\email{luyusheng@tongji.edu.cn}

\author{Zhaocheng Du}%\footnotemark[1]}
\authornotemark[1]
\affiliation{
\institution{Huawei Noah’s Ark Lab}
\state{Shenzhen}
\country{China}
}
\email{zhaochengdu@huawei.com}

\author{Xiangyang Li}
\affiliation{
\institution{Huawei Noah’s Ark Lab}
\state{Shenzhen}
\country{China}
}
\email{lixiangyang34@huawei.com}

\author{Pengyue Jia}
\affiliation{
\institution{City University of Hong Kong}
\state{Hong Kong}
%\state{Shenzhen}
\country{China}
}
\email{jia.pengyue@my.cityu.edu.hk}

\author{Yejing Wang}
\affiliation{
\institution{City University of Hong Kong}
\state{Hong Kong}
%\state{Shenzhen}
\country{China}
}
\email{yejing.wang@my.cityu.edu.hk}

\author{Weiwen Liu}
\affiliation{
\institution{Shanghai Jiao Tong University}
\state{Shanghai}
\country{China}
}
\email{wwliu@sjtu.edu.cn}

\author{Yichao Wang}
\affiliation{
\institution{Huawei Noah’s Ark Lab}
\state{Shenzhen}
\country{China}
}
\email{wangyichao5@huawei.com}

\author{Huifeng Guo}
\affiliation{
\institution{Huawei Noah’s Ark Lab}
\state{Shenzhen}
\country{China}
}
\email{huifeng.guo@huawei.com}

\author{Ruiming Tang}
\affiliation{
\institution{Huawei Noah’s Ark Lab}
\state{Shenzhen}
\country{China}
}
\email{tangruiming@huawei.com}

\author{Zhenhua Dong}
\affiliation{
\institution{Huawei Noah’s Ark Lab}
\state{Shenzhen}
\country{China}
}
\email{dongzhenhua@huawei.com}

\author{Yongrui Duan}%\footnotemark[2]}
\authornote{Corresponding authors.}
\affiliation{
\institution{Tongji University}
\state{Shanghai}
\country{China}
}
\email{yrduan@tongji.edu.cn}

\author{Xiangyu Zhao}%\footnotemark[2]}
\authornotemark[2]
\affiliation{
\institution{City University of Hong Kong}
\state{Hong Kong}
%\state{Shenzhen}
\country{China}
}
\email{xianzhao@cityu.edu.hk}

\renewcommand{\shortauthors}{Yusheng Lu et al.}
%% If needed use a foot or authornote to identify equal contribution

\begin{abstract}
Large Language Models (LLMs) have exhibited significant promise in recommender systems by empowering user profiles with their extensive world knowledge and superior reasoning capabilities. 
However, LLMs face challenges like unstable instruction compliance, modality gaps, and high inference latency, leading to textual noise and limiting their effectiveness in recommender systems. 
To address these challenges, we propose UserIP-Tuning, which uses prompt-tuning to infer user profiles. It integrates the causal relationship between user profiles and behavior sequences into LLMs' prompts. It employs Expectation Maximization (EM) to infer the embedded latent profile, minimizing textual noise by fixing the prompt template. 
Furthermore, a profile quantization codebook bridges the modality gap by categorizing profile embeddings into collaborative IDs pre-stored for online deployment. This improves time efficiency and reduces memory usage. 
Experiments show that UserIP-Tuning outperforms state-of-the-art recommendation algorithms. An industry application confirms its effectiveness, robustness, and transferability. 
The presented solution has been deployed in Huawei AppGallery’s Explore page since May 2025, serving \textbf{2 million daily active users}, delivering significant improvements in real-world recommendation scenarios.
The code is publicly available for replication at https://github.com/Applied-Machine-Learning-Lab/UserIP-Tuning.
\end{abstract}






\maketitle

\section{Introduction}
\label{sec:introduction}

\begin{figure}[!htbp]
    % \vspace{-2mm}
    \setlength\abovecaptionskip{0.3\baselineskip}
    \setlength\belowcaptionskip{0.2\baselineskip}
  \centering
    \includegraphics[width=0.95\linewidth]{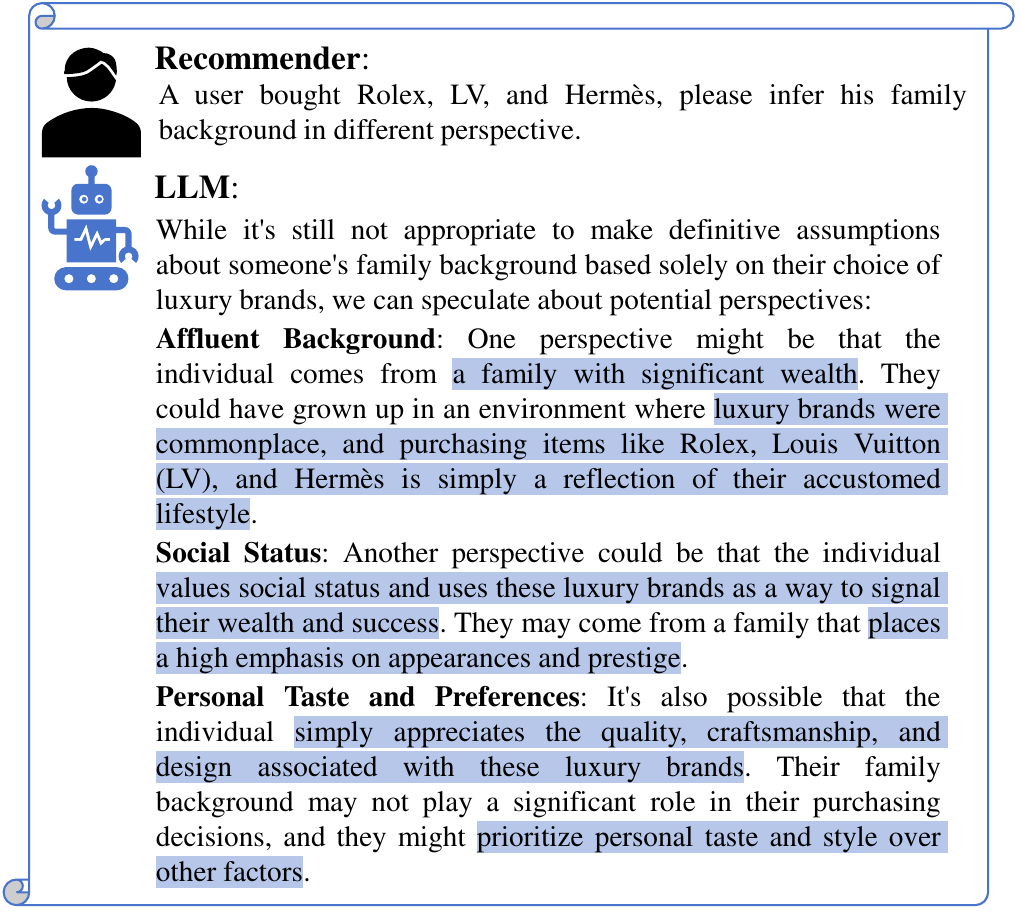}
  %\vspace{-0.1in}
  \caption{Example of inferring user latent profiles with LLMs based on observable behaviors. Blue lines indicate informative profiles, while white lines represent noise in the RS task.}
  \label{fig:intr}
  %\vspace{-0.2in}
  \vspace*{-6mm}
\end{figure}

Recommender Systems (RSs) suggest items based on users' historical records and preferences~\cite{zhao2018deeppage,
zhao2018deers,luo2023recranker,li2023e4srec,hou2023large}, alleviating information overload. However, RS often lacks users' latent profiles—users' underlying motivations behind behaviors—due to privacy concerns or these profiles' abstract nature~\cite{lin2023autodenoise,zhou2022infer,wu2020joint}. Family backgrounds and career information are frequently missing or hard to quantify. A survey~\cite{zhou2022infer} illustrated that only 40\% of Facebook users disclose career details. Without these profiles, recommendation features cannot accurately represent users and infer latent profiles.

With superior reasoning capabilities and extensive world knowledge \cite{li2023strec,liu2024leader,touvron2023llama1,touvron2023llama2,gautier2023JMLR,santis2024}, LLMs offer a promising solution for inferring user latent profiles based on observable behaviors.
For example, by analyzing user-purchased luxury items (Figure~\ref{fig:intr}), LLMs can infer quantifiable latent profiles like ``career success'' and ``social status'', which can be used as user features to improve performance.

\begin{figure}[!htbp]
    \setlength\abovecaptionskip{0.3\baselineskip}
    \setlength\belowcaptionskip{0.2\baselineskip}
  \centering
    \vspace{-0.1in}
    \includegraphics[width=0.85\linewidth]{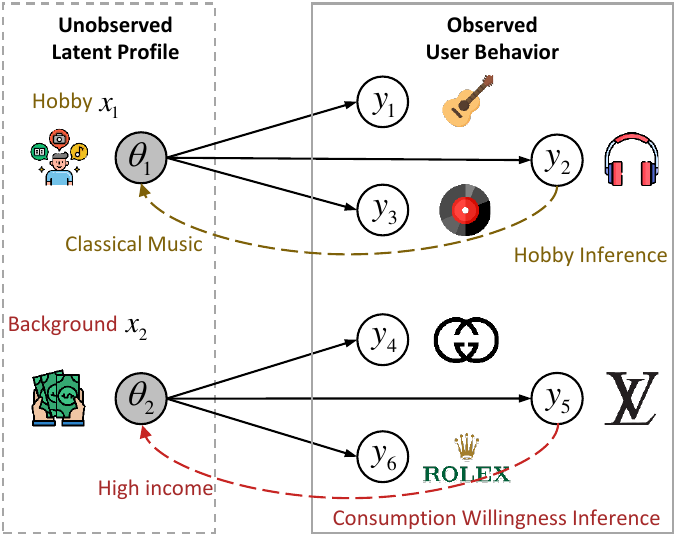}
  
  \caption{Users' latent profile and observed behavior.}
  \label{fig:unob}
  %\vspace{-2mm}
  %\vspace{-0.1in}
  \vspace{-6mm}
\end{figure}

Significant efforts have explored enhancing RS with LLMs' reasoning abilities~\cite{li2023hamur,zhang2024m3oe,zhang2023dpt,zhou2022infer,wang2023large}. KAR~\cite{xi2023towards} designs prompts to trigger profile inference, then embeds the inferred text with BERT~\cite{kenton2019bert} for deployment. LFM~\cite{wang2023large} generates text-based profiles from interaction histories to guide LLM-based RS tasks. Despite progress, these approaches face notable theoretical and practical challenges:  

\textbf{1) Twisted Causality: }  
LLMs generate tokens autoregressively~\cite{wang2025crossdistillation,wang2024plate,lester2021power,xie2022explanation}, while in RS, user profiles are causal factors of interactions (Figure~\ref{fig:unob}). Many prompt-based methods reverse this order, describing behaviors first and deriving profiles afterward. Such inversion may misalign causal semantics and limit generalization. 
\textbf{2) Textual Noise: }  
Due to unstable instruction following~\cite{fu2025unified,taori2023stanford,touvron2023llama2,dubois2024alpacafarm} and CoT reasoning~\cite{wei2022chain}, inferred profiles often contain noise~\cite{yang2023cptpp,cai2023low,tan2023evaluation,gao2024smlp4rec} (Figure~\ref{fig:intr}). Incorporating tunable prompts can help concentrate information and suppress irrelevant text. 
\textbf{3) Modality Gap: }  
The complex syntax and semantics in LLM outputs are difficult for RS to leverage effectively~\cite{zhang2023collm,tallrec23,yu2025uniform,luo2025tapo,liu2025bridge}. RS struggles to extract useful collaborative signals from such embeddings. 
\textbf{4) Reasoning Inefficiency: }  
Billion-parameter LLMs challenge RS latency constraints~\cite{wang2025pre,gao2025llm4rerank,gao2025samplellm,wang2024rethinking}, while large embeddings further raise computational costs. Mapping them into compact, sparse features (e.g., ``gender'', ``race'') may improve efficiency.

Motivated by the outlined challenges and inspired by recent advancements in soft-prompt tuning techniques~\cite{liu2025contrastive,zhang2025llm,liu2025llmemb,lester2021power,li2023pepler,zhang2023promptst}, we propose the \textbf{User} \textbf{I}nherent \textbf{P}rofile inference with prompt \textbf{Tuning} (UserIP-Tuning) framework, which is lightweight, controllable, and easily integrated with any recommendation model. It consists of a UserIP inference module, a UserIP quantization module, and a pre-stored UserIP feature bank.

% To address the first and the second challenges, the UserIP inference module treats the user's inherent profiles as trainable soft tokens embedded in a prompt template. The prompt is manually designed to stimulate LLMs to generate ground-truth user interaction behaviors sequence. These soft tokens are then inferred with the EM algorithm to maximize the likelihood of this sequence under the distribution defined by a frozen LLMs.
% Then, to tackle the third and fourth issues, the UserIP quantization module maps these trained soft tokens into sparse feature IDs with a trainable codebook. The same ID shares similar semantic information, much like the classical sparse feature. Finally, these sparse IDs are pre-stored in a latent feature bank used for online serving to improve performance further. Experiments on four datasets demonstrate the effectiveness of UserIP-Tuning. The major contributions are summarized as follows: 
To address the first two challenges, the UserIP inference module treats user latent profiles as trainable soft tokens within a prompt template. This prompt elicits accurate user behavior sequences from LLMs. The soft tokens are inferred using the EM algorithm, which maximizes the likelihood of these sequences based on a frozen LLM distribution.
For the remaining two challenges, the UserIP quantization module converts trained soft tokens into sparse feature IDs using trainable codebooks. These IDs, which capture both semantic similarity and collaborative signals, function like traditional sparse features and are stored in a latent feature bank for online deployment. The contributions are as follows:
\vspace{-0.02in}
\begin{itemize}[leftmargin=*]
% \item We propose a UserIP-Tuning, a light, efficient, and model-agnostic user inherent profile inference framework that can further improve recommendation models' performance with guaranteed inference efficiency.
\item UserIP inference module is the first LLM soft-prompt-based user profile inference algorithm that can improve profile inference's causality and avoid textual noises.  
% \item The proposed UserIP inference module is the first soft prompt tuning based user-profile inferring algorithm. The inferred profiles are more explainable, noise-mitigated, and causality-sensitive compared to LLM generated profiles.
\item The framework is efficient and model-agnostic and can improve the RS's performance with guaranteed inference efficiency.
% \item The proposed UserIP quantization module builds a bridge between the LLMs' semantics modality and the recommendation modality by converting language semantics token into sparse and pre-storable id features. This module significantly boosts the inference efficiency and accuracy of the LLM-assisted recommendation model;

% \item We conducted extensive experiments to demonstrate the advantages of UserIP-tuning on public and industrial datasets in different perspectives like effectiveness, efficiency, generalizability, and explainability.
\item Extensive experiments on public and industrial datasets, and online A/B testing, validate the effectiveness, efficiency, generalizability, and explainability of UserIP-Tuning.
\end{itemize}

\section{Framework}
\label{sec:framework}

\begin{figure*}[t]
        \setlength\abovecaptionskip{0.3\baselineskip}
        \setlength\belowcaptionskip{0.1\baselineskip}
	\centering
	%	\hspace*{-7mm}1.049
	\includegraphics[width=0.95\linewidth]{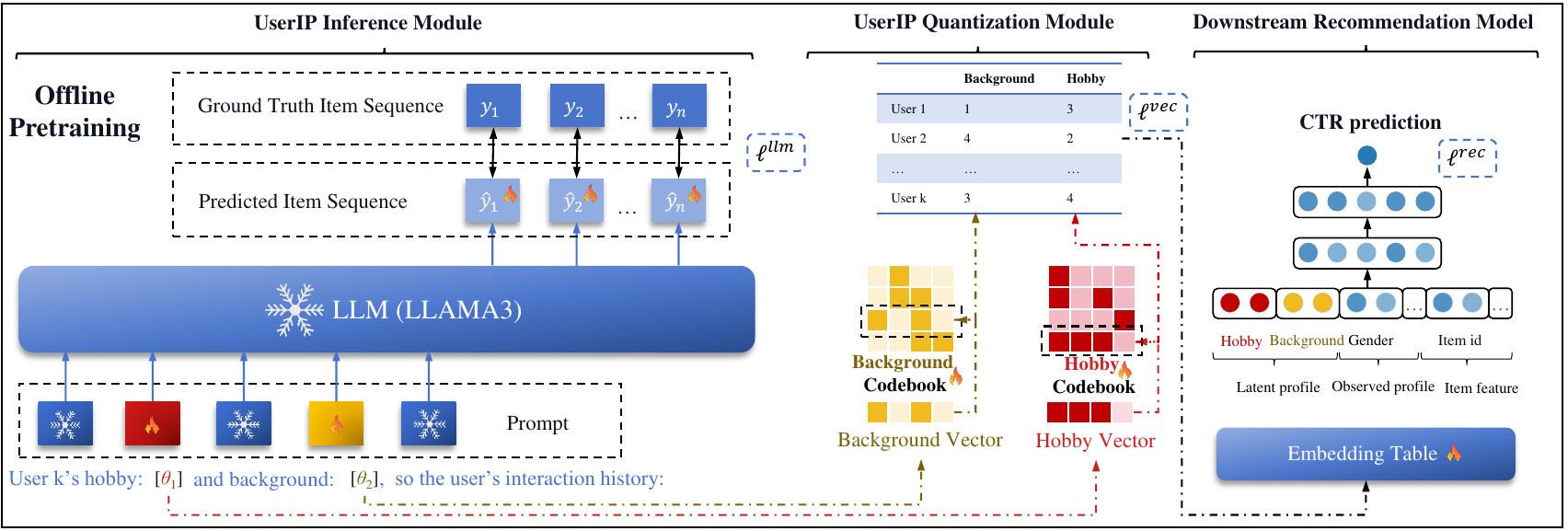}
	\caption{Overview of the UserIP-Tuning framework. Here, two user latent profiles are illustrated: hobby and income background. UserIP-Tuning consists of a UserIP inference module, a UserIP quantization module, and a pre-stored UserIP feature bank.}
	\label{fig:Fig1_Overview}
	\vspace{-4mm}
\end{figure*}

% This section presents the technical details of the UserIP-Tuning framework. First, the problem setting and preliminary are described.

\subsection{Preliminary and Setup}
\label{subsec:pre}
A recommendation model can be formulated as learning a mapping $f$ from a feature set $E$ to the target $y^{rec}$: $y^{rec} = f(E)$. The feature set $E$ includes observable attributes $E_{obs}$ (e.g., gender, age) derived from logs, and unobservable latent profiles $E_{latent}$ (e.g., family background, personal traits) that are often missing due to privacy concerns or an abstract nature.  

Similar to $E_{obs}$, $E_{latent}$ also drives user behaviors, as shown in Figure~\ref{fig:unob}. For instance, users from wealthy families are more likely to purchase luxury goods, while music lovers may collect instruments. We represent $E_{latent}$ as a set of trainable soft tokens ${\theta^1, \theta^2, ..., \theta^M}$, where each token corresponds to a latent profile variable and is learnable. The task then reduces to inferring the optimal latent variables that maximize the likelihood of observed user behaviors.  
%\vspace{-5pt} % 减少上方间距
\begin{equation}
\setlength\abovedisplayskip{6pt}%shrink space
\setlength\belowdisplayskip{6pt}
\small
%\vspace{-2mm}
E_{latent} = argmax_\theta P_\phi(y^{rec}|\theta^1, \theta^2, ..., \theta^M).
\label{inference}
%\vspace{-2mm}
\end{equation}
%\vspace{-2pt} % 减少上方间距
where $P_\phi$ maps latent profiles to the probability of observed behaviors. However, such a function is unavailable in RS since $\theta^1, \theta^2, ..., \theta^M$ are unobserved. LLMs, with their rich knowledge and reasoning ability, can act as a surrogate for $P_\phi$. UserIP-Tuning leverages this to infer latent profiles in semantic space while simultaneously addressing the four challenges outlined in the Introduction.  

%\vspace{-0.1in}
\subsection{Framework Overview}
The process of UserIP-Tuning is illustrated in  Figure~\ref{fig:Fig1_Overview}. 
%It consists of a UserIP inference module, a UserIP quantization module, and a pre-stored UserIP feature bank. 
The inference module treats users' latent profiles as soft prompts and infers them by maximizing the likelihood of the ground-truth user interaction sequence. Then, these latent profiles are hashed into discrete collaborative IDs by a trainable quantization module supervised by recommendation loss. Finally, these IDs are stored in the offline feature bank for downstream RS tasks.

%\vspace{-0.1in}
\subsection{UserIP Inference Module}
To address the twisted causal relationship and the textual noise issues mentioned in the Introduction, the UserIP inference module models the latent profile inference problem (Figure \ref{fig:unob}) in the LLMs' semantic space with task-specific soft prompts and causal masks. 

\begin{figure}[h]
    \setlength\abovecaptionskip{0.2\baselineskip}
    \setlength\belowcaptionskip{0.1\baselineskip}
  \centering
    \includegraphics[width=0.7\linewidth]{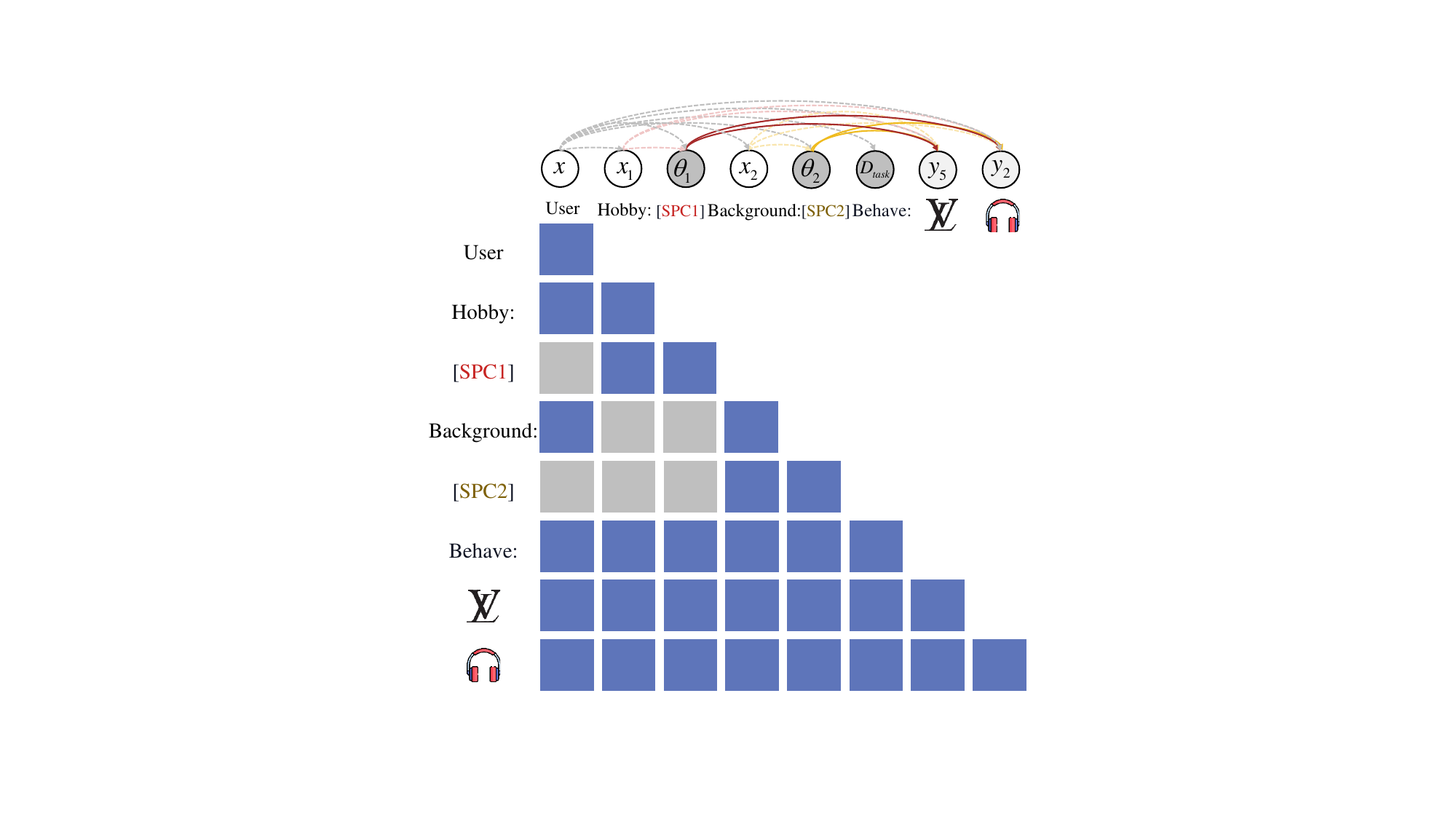}
  \caption{Causal relationship (Upper) between missing user profiles and behaviors. The curve means the causal direction. Note that latent profiles are independent of each other. Causal mask (Below) in UserIP Inference Module. The blue (gray) square denotes that column $j$ will (not) attend to row $i$.}
  \label{fig:pr}
  \vspace{-4mm}
\end{figure}

Firstly, the probability graph demonstrated in Figure~\ref{fig:unob} is reorganized to adapt to the LLMs' autoregressive structure as shown in Figure~\ref{fig:pr} (Up), where the left part is treated as the prompt (the fixed hobby, background profiles) and the right part is treated as the ground-truth output (Behave: LV, over-ear headphones). To reduce the textual noise, only tokens ($\theta$) next to target latent profiles' names ($x$) are treated as soft prompts. These soft prompts can only be influenced by their respective latent profiles' names.

A text prompt is then designed to satisfy the above causal relationship and consists of three components. (1) Context tokens $x$ to indicate the latent profiles we are interested in, (2) soft latent profile tokens $\theta$ that can be tuned to maximize the likelihood of user interaction behaviors, and (3) a specific task description $D_{task}$ such as user click history prediction or review prediction for a target item. The template of the prompt is presented:
%\vspace{-5pt} % 减少上方间距
\begin{equation} 
\setlength\abovedisplayskip{6pt}%shrink space
\setlength\belowdisplayskip{6pt}
\small
[L_{M}^{'}, D_{task} ]=[x^1, \theta^1,x^2, \theta^2,..., x^M, \theta^M, D_{task}]
\end{equation}
%\vspace{-1pt} % 减少上方间距
where $L_{M}^{'}$ denotes the concatenated latent profile names $x$ and soft latent profile tokens $\theta$. For instance, Figure~\ref{fig:Fig1_Overview} and the following prompt template illustrate the prompt where $x^1$ is ``This user's hobby is'' and $x^2$ is ``his background is''. By putting each profile name before its corresponding latent profile token, LLMs are forced to infer latent profiles more controllably because LLMs are trained to consider language context in NLP tasks. The backward inferred $\theta^m$ means the accurate representation of the latent profile $x^m$.

\begin{figure}[h]
 \setlength\abovecaptionskip{-0.01\baselineskip}
 \setlength\belowcaptionskip{-0.1\baselineskip}
    \vspace{-3mm}
    \centering
    \includegraphics[width=0.99\linewidth]{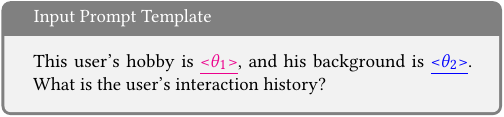}
    \caption{Prompt template of user's latent profiles}
    \label{fig:tcolor}
    \vspace{-3mm}
\end{figure}
% \begin{tcolorbox}[%`colback`=gray!10,
%             colframe=gray,
%             width=1\linewidth,
%             arc=1mm, 
%             auto outer arc,
%             title={Input Prompt Template},
%             breakable,]
            
%             This user's hobby is \textcolor{magenta}{\underline{<$\theta_1$>}}, and his background is \textcolor{blue}{\underline{<$\theta_2$>}}. What is the user's interaction history?
% \end{tcolorbox} % 在arxiv里面编译有问题，文字重叠

To restrict soft prompts to their corresponding latent profiles, we apply causal masks during training (Figure~\ref{fig:pr}, Down). This mask ensures that each latent profile variable $\theta^1$ and $\theta^2$ only attends to its own profile name and interactions, ignoring unrelated words.

Given the prompt and causal mask, the LLM generates a probability distribution over user interaction sequences. We maximize the likelihood of the ground-truth sequence via gradient descent, updating only the soft token embedding of the $m$-th latent profile $\theta^m$ while keeping other LLM parameters fixed. Each embedding reflects a profile type (e.g., high income in income status), which helps capture user preferences.

LLMs aim to predict the user $u$'s interaction history while updating the latent profile variables $\theta^m$. We build the language modeling losses (Cross-Entropy losses) leveraging the ground-truth user interaction history $y^{rec}$ and the generated result $\hat{y}^{rec}$. Specifically, after inferring multiple latent profile variables, the fine-tuning objective is to minimize $ \mathcal{L}(\theta)= \mathbb{E}_{u}[\ell_{u}(x^1, x^2, ..., x^M,y^{rec},\theta^1, \theta^2, ..., \theta^M)] $:
\begin{equation}
\setlength\abovedisplayskip{6pt}%shrink space
\setlength\belowdisplayskip{6pt}
\small
\label{equ:LMloss}
\begin{aligned}
    & \ell^{llm}_{u} (x^1, x^2, ..., x^M,y^{rec},\theta^1, \theta^2, ..., \theta^M) \\
    & = -log P_{LLM}(y^{rec} | x^1, x^2, ..., x^M ,\theta^1, \theta^2, ..., \theta^M ) \\
\end{aligned}
\end{equation}

Then, the learned latent profile variables are input into the UserIP quantization module to eliminate the out-of-range noises. %Some theoretical analysis is in Appendix~\ref{app:1}.

%\vspace{-0.1in}
\subsection{UserIP Quantization Module}
Latent profile embeddings inferred above lie in the semantic space of LLMs. Although rich in semantics, they are overly detailed for RS, which mainly relies on collaborative signals. The high dimensionality of LLM embeddings also hinders efficient online inference.  

To address this modality gap between the NLP tasks and RS tasks and improve efficiency, we design a quantization module that extracts lightweight collaborative IDs from latent profile embeddings. For each inferred latent profile, a codebook $\mathcal{C}^m$ with size $K$ is assigned, and the $k$-th code embedding is denoted as $\{\boldsymbol{v}^{m}_{k}\}$.  

Unlike standard CV/NLP quantization methods~\cite{van2017vqvae,rajput2023recommender,zheng2023adapting} that rely only on semantic distance, UserIP incorporates collaborative signals as supervision. Since RS prioritizes collaborative modeling, embeddings that differ semantically but align in collaborative space (e.g., the Beer \& Diapers case) are considered similar.  
  
\par
To integrate the collaborative objective, we design the following loss function (Equation \eqref{equ:VQloss}). %The first term of 
The loss function drags the codebook close to the latent profile space. $sg[\cdot]$ denotes the stop-gradient operator and $\beta$ is loss weight parameters. 
The first part minimizes the distance between the cluster centroid vectors $v_k^m$ and the latent profile variables $\theta^m$. The second part constrains the profile embedding's updating speed.
%The second term trains the latent profile embeddings with a surrogate recommendation model loss. $v_{c_m}^m$ denotes the selected codebook embedding (Equation \eqref{equ:nearest}), $i$ stands for item embedding and $E_{obs}$ stands for other user profiles. It directly uses codebook embeddings as the input of recommendation tasks and uses the recommendation loss update latent profile embedding (together with $\ell^{llm}$)  by straight-through gradient estimation. 
%\vspace{-5pt}
\begin{equation}
\setlength\abovedisplayskip{6pt}%shrink space
\setlength\belowdisplayskip{6pt}
\small
\label{equ:VQloss}
\begin{aligned}
    \ell^{vec}_{u} =& \sum_{m=1}^{M} [\left \| sg[\theta^m] - v_k^m \right \|^2_2 + \beta \left \| \theta^m - sg[v_k^m] \right \|^2_2] %\\
    %& + BCE(y^{rec}, Rec^s(v_{c_1}^1,...v_{c_m}^m, E_{obs},i_u)) 
\end{aligned}
\end{equation}
%\vspace{-2pt}
After training, a user's latent profile ID is inferred by finding the index of the codebook embedding closest to the latent profile inferred from the previous section.
%\vspace{-7pt}
\begin{equation}
\setlength\abovedisplayskip{6pt}%shrink space
\setlength\belowdisplayskip{6pt}
\small
\label{equ:nearest}
    c_m=arg \min_{k} \left \| \theta^m - v_k^m \right \| ^2_2
\end{equation}
%\vspace{-1pt}
Thus, $v_{c_m}^m$ denotes the selected codebook embedding. The loss for latent profile training and quantization is defined as follows, 
%\vspace{-5pt}
\begin{equation}
\setlength\abovedisplayskip{6pt}%shrink space
\setlength\belowdisplayskip{6pt}
\small
     \mathcal{L} = \ell^{llm} + \alpha * \ell^{vec} %+ \gamma * \ell^{rec}
\end{equation}
%\vspace{-1pt}
where $\alpha$ is the loss weight.
%\vspace{-1mm}
%\vspace{-0.1in}
\subsection{UserIP Feature Bank and Downstream Recommender model}
Training large text embeddings and soft tokens substantially increases computational overhead, leading to inefficiencies in online RS reasoning.
To solve the reasoning inefficiency problem, after training UserIP-Tuning, a feature bank will be used to store the latent profile indices $c_1$, ..., $c_M$ of each user. 
The users' latent profile indices are added into the traditional RS, such as DCN \cite{wang2017dcn}, to augment user modeling features and improve collaborative signals.  For a given user-item pair, the recommendation prediction outcomes are $\hat{y}^{rec} = Rec[u,i,c_1,c_2,...,c_M,E_{obs}]$. The single target loss function is defined as: $\ell^{rec}_{u,i}=-\sum_{n=1}^{N} y^{rec}_n log (\hat{y}^{rec}_n)$ 
where $Rec[\cdot\cdot\cdot]$ represents a traditional recommendation model, $E_{obs}$ stands for other observable user profiles.

The total loss trains the latent profile embeddings with a surrogate recommendation model loss. 
%It directly uses codebook embeddings indices as the input of RS tasks and 
It directly uses the recommendation loss to update latent profile embedding (together with $\ell^{llm}$, $\ell^{vec}$)  by straight-through gradient estimation. 
%\vspace{-7pt}
\begin{equation}
\setlength\abovedisplayskip{6pt}%shrink space
\setlength\belowdisplayskip{6pt}
\small
     \mathcal{L} = \ell^{llm} + \alpha * \ell^{vec} + \gamma * \ell^{rec}
\end{equation}
%\vspace{-2pt}
where $\gamma$ is the loss weight. %Our algorithm process is summarized in Appendix~\ref{app:2}.
The optimization process is formulated in the Algorithm ~\ref{alg:UserIP-Tuning}.
\begin{algorithm}[]
	\caption{\label{alg:UserIP-Tuning} An Algorithm for UserIP-Tuning.}
	\raggedright
	{\bf Input}: Samples in dataset $\mathcal{D}=\{(x^1, x^2, ..., x^M, D_{task}, y^{rec})\}_{u,i}$ for different user-item pairs $(u,i)$, LLMs Llama3, user profiles tokens $\theta^1$, $\theta^2$, ..., $\theta^M$, codebook sizes for profiles $K_1,K_2,...,K_M$. Freeze all parameters in Llama3 except $\theta^m$.\\
	{\bf Output}: Fine-tuned user latent profiles embedding, user profiles indices, quantization codebooks. \\
	\begin{algorithmic} [1]
		\WHILE{not converged}
		\STATE Sample a random batch $\mathcal{B}$ in $\mathcal{D}$ and initialize gradients $g_1 \gets 0$ and $g_2 \gets 0$, and cluster centroids $\boldsymbol{v}^{m}_{k}$
		\FOR{each data point $(x^1, x^2, ..., x^M, D_{task_i}, y^{rec})$ in $\mathcal{B}$}
            \STATE Update gradient $g_1=g_1+\frac{\partial \ell^{llm}_{u,i}(x^1, x^2, ..., x^M,y^{rec}\theta^1, \theta^2, ..., \theta^M)}{\partial \textit{E}_{latent}} $ based on Eq. \ref{equ:LMloss}.
            \STATE Calculate the nearest neighbor indices for profiles $c_1$, ..., $c_M$ via Eq. \ref{equ:nearest} and update gradient $g_2=g_2+\frac{\partial \ell^{vec}_{u,i}(\textit{E}_{latent},v_k^m)}{\partial \boldsymbol{v}^{m}_{k} } $ based on Eq. \ref{equ:VQloss}.
            %\STATE Add user profiles indices $c_1$, ..., $c_M$ for downstream recommender to calculate single target loss $\ell^{rec}_{u,i}$ and total loss.
            \STATE Update the latent variables $\textit{E}_{latent}=\textit{E}_{latent} - \alpha (g_1+g_2) $.
            \STATE Update the quantization codebook $\boldsymbol{v}^{m}_{k}=\boldsymbol{v}^{m}_{k} - \alpha g_2$.
            \ENDFOR
		\ENDWHILE
            \RETURN Latent profile indices for downstream recommender.
	\end{algorithmic}
    
\end{algorithm}

\vspace{-0.1in}
\section{Experiment}
\label{sec:experiment}
%\vspace{-0.05in}
\subsection{Datasets, Evaluation Metrics, and Baselines}
\label{sec:Datasets}
Our model is evaluated on four real-world open datasets, including Amazon\footnote{\url{https://cseweb.ucsd.edu/~jmcauley/datasets/amazon/links.html}} Clothing Shoes and Jewelry, Movies and TV, Video Games, and Yelp\footnote{https://www.yelp.com/dataset/documentation/main}. For simplicity, we use Clothing, Movies, and Games to denote the first three datasets. Below is the general information about the datasets in Table~\ref{table:statistics}. 
\begin{table}[!htbp]
\vspace{-1mm}
        \setlength{\abovecaptionskip}{0cm} %# 调整间距
	\setlength{\belowcaptionskip}{0cm}
	\small
	\caption{Statistics of the datasets.}
	\label{table:statistics}
	\begin{tabular}{@{} ccccc @{}}
		\toprule[1pt]
		Data & Clothing & Movies & Games & Yelp \\ \midrule
	 Interactions & 179,223 & 441,783 & 19,927 & 1,293,247 \\
		 Users & 38,764 & 7,506 & 2,490 & 27,147 \\
		 Items & 22,919 & 7,360 & 8,912 & 20,266 \\
		 Sparsity($\%$) & 99.98 & 99.20 & 99.91 & 99.76 \\ % & rating 1$\sim$5 \\ 
            \bottomrule[1pt]
	\end{tabular}
	\vspace{-3mm}
\end{table}

To verify recommendation effectiveness, we compute the Area Under Curve (AUC), Logloss. Following \cite{li2023ctrl, yan2014coupled}, the relative improvements (\textit{R.I.}) of UserIP-Tuning's AUC and Logloss compared with the best baseline are calculated as follows:
\begin{equation}
\small
\begin{aligned}
    & R. I. of AUC = (\frac{AUC(model)-0.5}{AUC(baseline)-0.5} - 1) \\
\end{aligned}
\end{equation}
\begin{equation}
\small
\begin{aligned}
     &R. I. of Logloss = \frac{Logloss(baseline)-Logloss(model)} {Logloss(model)} 
\end{aligned}
\end{equation}

We use three types of RS baselines: shallow algorithms, advanced deep learning algorithms, and LLM-based RS methods.
\begin{table*}[!htbp]
     \setlength{\abovecaptionskip}{0cm} %# 调整间距
	 \setlength{\belowcaptionskip}{-0.01cm}
	\small
	\caption{Performance comparison of different baseline methods.}
	\label{table:result1}
        \resizebox{1\textwidth}{!}{
	\begin{tabular*}{1\linewidth}{@{\extracolsep{\fill}} p{0.5cm}cp{0.9cm}p{0.9cm}p{0.9cm}p{0.9cm}p{0.9cm}p{0.9cm}p{0.9cm}p{0.9cm} }
		\toprule[1pt]
		\multirow{2}{*}{Types} & \multirow{2}{*}{Methods} & \multicolumn{2}{c}{Clothing} & \multicolumn{2}{c}{Movies} & \multicolumn{2}{c}{Games} & \multicolumn{2}{c}{Yelp} \\ \cmidrule(l){3-10} 
		& & AUC & Logloss & AUC & Logloss & AUC & Logloss & AUC & Logloss \\ \midrule 
		\multirow{2}{*}{Shallow} & FFM & 0.5588 & 0.6212 & 0.7998 & 0.5090 & 0.6946 & 0.8347 & 0.6931 & 0.6903 \\
		& AFM & 0.5524 & 0.6181 & 0.7876 & 0.5656 & 0.6627 & 1.0342 & 0.6955 & 0.6836 \\ \midrule
		\multirow{9}{*}{Deep} & FiBiNet & 0.5952 & 0.6198 & 0.8111 & 0.4659 & 0.6862 & 0.8823 & 0.7243 & 0.5621 \\ 
		& DIFM & 0.5935 & 0.5758 & 0.8054 & 0.4694 & 0.6897 & 0.9285 & 0.7174 & 0.5666 \\ 
            & AFN & 0.5945 & 0.5835 & 0.8050 & 0.4796 & 0.7022 & 0.8052 & 0.7169 & 0.6008 \\
            & DeepFM & 0.6015 & 0.5965 & 0.8061 & 0.4729 & 0.7221 & 0.8659 & 0.7017 & 0.6028 \\
            & AutoInt & 0.6023 & 0.6023 & 0.8072 & 0.4842 & 0.7257 & 0.8205 & 0.7194 & 0.6122 \\
            & ONN & 0.6094 & 0.5903 & 0.8121 & 0.4598 & 0.7223 & 0.8658 & 0.7253 & 0.5598 \\
            & PNN & 0.6088 & 0.5874 & 0.8154 & 0.4570 & 0.7178 & 0.7874 & 0.7287 & 0.5574 \\
            & DCNv2 & 0.6148 & 0.5848 & 0.8134 & 0.4643 & 0.7263 & 0.7991 & 0.7237 & 0.5704 \\
            & DCN & 0.6214 & 0.5322 & 0.8167 & 0.4582 & 0.7317 & 0.7744 & 0.7259 & 0.5590 \\ \midrule
            \multirow{5}{*}{LLMs}& KAR & 0.6003 &	0.5925	& 0.8058 & 0.4885	& 0.7136 & 0.8679 & 0.7146 & 0.5849 \\
            & PEPLER & 0.6101 & 0.5726 & 0.8095 & 0.4726 & 0.7188 & 0.8347 & 0.7227 & 0.5784\\
            & ReLLa & 0.6214 & 0.5320 & 0.8099 & 0.4721 & 0.7322 & 0.7814 & 0.7290 & 0.5587\\
            % & UserIP-Tuning-gpt2 & 0.6234 & 0.5319 & 0.8179 & 0.4571 & 0.7329 & 0.7801 & 0.7311 & 0.5562 \\
            & UserIP-Tuning-Llama2 & \ul{0.6269} & \ul{0.5126} & \ul{0.8184} & \ul{0.4566} & \ul{0.7393} & \ul{0.7285} & \ul{0.7314} & \ul{0.5558} \\
            & UserIP-Tuning-Llama3 & \textbf{0.6272*} & \textbf{0.5033*} & \textbf{0.8189*} & \textbf{0.4517*} & \textbf{0.7398*} & \textbf{0.7225*} & \textbf{0.7320*} & \textbf{0.5502*} \\ \midrule
            \multicolumn{2}{c}{\textit{R.I.}} & 4.7690\% & 5.7342\% & 0.6819\% & 1.4368\% & 3.4870\% & 7.1834\% & 1.4429\% & 1.3086\% \\\bottomrule[1pt]
	\end{tabular*}
    }
        {{\small  The symbol * denotes the significance level with $p \leq 0.05$. \textbf{Bold} font indicates the best-performing method.}}
        \vspace{-2mm}
\end{table*}

\vspace{-0.1in}
\subsection{Implementation Details}
\label{app:imple}
We select Llama3-8B, Llama2-7B as the LLMs backbone in UserIP-Tuning. We implement all the compared methods using Python 3.9 and PyTorch 2.1.0. Following previous work~\cite{li2023pepler,liao2023llara}, the training, validation, and test sets are divided into 8:1:1. The benchmark hyperparameters are set by default to obtain their optimal performance, and the Adam optimizer is used. The embedding size in Llama3-8B and Llama2-7B is 4096, and the batch size is 8. We optimize UserIP-Tuning with AdamW. During training, we freeze the LLM's weights. The learning rate is 0.001, and the latent profile embedding size is 4096. Moreover, the number of hobby codebook is set to 4, and the number of background codebook is 3. In the downstream recommender DCN, the number of network layers is 3, the embedding dimension is 8, the dropout rate is 0.2, and the MLP embedding dimensions are (16, 16). We use item ID and user ID as the obvious feature fields. The loss weights are $\alpha$=0.001, $\beta$=0.001, $\gamma$=0.001. 
% We use one NVIDIA H800 80GB GPU.

\vspace{-0.1in}
\subsection{Overall Performance Comparison}
\label{sec:RQ1}
We compare UserIP-Tuning with advanced benchmarks, with results summarized in Table~\ref{table:result1}. The best scores are highlighted in bold, and the second best are underlined.  

\textit{\textbf{1)}} UserIP-Tuning consistently outperforms all baselines. Its \textit{R.I.} gains in AUC are 4.77\%, 0.68\%, 3.49\%, and 1.44\% on Clothing, Movies, Games, and Yelp, respectively. For Logloss, the improvements are 5.73\%, 1.44\%, 7.18\%, and 1.31\%, respectively. The larger relative gains in Logloss show enhanced precision in modeling user preferences.  
\textit{\textbf{2)}} Shallow benchmarks yield the poorest results, relying only on user-id and item-id features, which limit recommendation effectiveness.  
\textit{\textbf{3)}} Deep learning baselines perform better by capturing diverse and collaborative signals, yet remain inferior to UserIP-Tuning, which further incorporates user profile variables and collaborative indices.  
\textit{\textbf{4)}} LLM-based methods surpass most deep models by leveraging reasoning over user preferences and factual item knowledge. Still, UserIP-Tuning achieves superior predictive performance by explicitly learning users’ inherent profiles.

\vspace{-0.1in}
\subsection{Transferability Study}
\label{app:trans}
We investigate the generalization transferability and apply UserIP-Tuning to other downstream models. The results are demonstrated in Table~\ref{table:transferability}. We find UserIP-Tuning + PNN ( ONN or DCNv2) performs superior to the original PNN (ONN or DCNv2), which indicates that the user latent profiles variables and corresponding indices can strengthen multiple recommenders' performance.
% This part investigates the generalization transferability of UserIP-Tuning. Specifically, we study whether the user profiles indices trained with UserIP-Tuning can be applied to other different downstream models $Rec[\cdot\cdot\cdot]$. 
% The user profiles indices are leveraged to train other RS, such as PNN et al. The results are demonstrated in Table~\ref{table:transferability}, where ``UserIP-Tuning-PNN'' represents the corresponding indices features in UserIP-Tuning are added to ``PNN''; similarly, the performance of ``UserIP-Tuning-ONN'' and ``UserIP-Tuning-DCNv2'' is evaluated. We see that UserIP-Tuning with PNN ( ONN or DCNv2) performs superior to the original PNN (ONN or DCNv2). This phenomenon indicates that the user latent profiles variables and corresponding indices can strengthen multiple recommenders' performance.
\begin{table*}[!htbp]
        \setlength{\abovecaptionskip}{0cm} %# 调整间距
	\setlength{\belowcaptionskip}{-0.1cm}
	\small
        \centering
	\caption{Transferability Study from DCN to other recommendation model.}
	\label{table:transferability}
\begin{tabular*}{1\linewidth}{@{\extracolsep{\fill}}ccp{0.9cm}p{0.9cm}p{0.9cm}p{0.9cm}p{0.9cm}p{0.9cm}p{0.9cm}p{0.9cm}}
\toprule[1pt]
\multirow{2}{*}{Functionality}      & \multirow{2}{*}{Models} & \multicolumn{2}{c}{Clothing} & \multicolumn{2}{c}{Movies} & \multicolumn{2}{c}{Games} & \multicolumn{2}{c}{Yelp} \\ \cline{3-10} 
&                         & AUC         & Logloss        & AUC        & Logloss       & AUC       & Logloss       & AUC       & Logloss      \\  \hline
Training Surrogate & UserIP-DCN      & 0.6272 & 0.5033 & 0.8189 & 0.4517 & 0.7398 & 0.7225 & 0.7320 & 0.5502 \\ \midrule[1pt]
\multirow{2}{*}{Transfer to PNN}   & PNN            & 0.6088 & 0.5874 & 0.8154 & 0.4570 & 0.7178 & 0.7874 & 0.7287 & 0.5574 \\
& UserIP-PNN     & \textbf{0.6137} & \textbf{0.5383} & \textbf{0.8159} & \textbf{0.4566} & \textbf{0.7208} & \textbf{0.8076} & \textbf{0.7284} & \textbf{0.5578}           \\  \midrule[1pt]
\multirow{2}{*}{Transfer to ONN}   & ONN            & 0.6094 & 0.5903 & 0.8121 & 0.4598 & 0.7223 & 0.8658 & 0.7253 & 0.5598 \\
& UserIP-ONN     & \textbf{0.6140} & \textbf{0.5021} & \textbf{0.8098} & \textbf{0.4626} & \textbf{0.7289} & \textbf{0.8395} & \textbf{0.7246} & \textbf{0.5613}   \\  \midrule[1pt]
\multirow{2}{*}{Transfer to DCNv2} & DCNv2         & 0.6148 & 0.5848 & 0.8134 & 0.4643 & 0.7263 & 0.7991 & 0.7237 & 0.5704 \\
& UserIP-DCNv2  & \textbf{0.6263} & \textbf{0.5125} & \textbf{0.8180} & \textbf{0.4544} & \textbf{0.7324} & \textbf{0.8625} & \textbf{0.7308} & \textbf{0.5569} \\ \bottomrule[1pt]           
\end{tabular*}
\vspace{-2mm}
\end{table*}
% \begin{center}
% \begin{table*}[!htbp]
%         \setlength{\abovecaptionskip}{0cm} %# 调整间距
% 	\setlength{\belowcaptionskip}{-0.1cm}
% 	\small
%         \centering
% 	\caption{Ablation study of the VQ module on four datasets.}
% 	\label{table:Abla}
%         \centerline{
% 	\begin{tabular}{@{} ccccccccc @{}}
% 		\toprule[1pt]
% 		\multirow{2}{*}{Model component} &  \multicolumn{2}{c}{Clothing} & \multicolumn{2}{c}{Movies} & \multicolumn{2}{c}{Games} & \multicolumn{2}{c}{Yelp} \\ \cmidrule(l){2-9} 
% 		& AUC & Logloss & AUC & Logloss & AUC & Logloss & AUC & Logloss  \\ \midrule
%             UserIP-Tuning w.o. VQ & 0.6201 & 0.5577 & 0.7966 & 0.4636 & 0.7340 & 0.8429 & 0.7277 & 0.5511 \\
%             UserIP-Tuning & \textbf{0.6269} & \textbf{0.5126} & \textbf{0.8184} & \textbf{0.4566} & \textbf{0.7393} & \textbf{0.7285} & \textbf{0.7314} & \textbf{0.5558}\\\bottomrule[1pt]
% 	\end{tabular}
% 	%	\vspace{-3mm}
%         }
% \end{table*}
% \end{center}

\vspace{-0.1in}
\subsection{Ablation Studies}
\label{sec:RQ3}

This subsection examines the ablation studies of the quantization module. The results are verified in Table~\ref{table:Abla}. When we replace the quantization module with K-Means and Agglomerate Features (A.F.) cluster methods or the quantization module is removed (w.o. quantization), the latent profile variables $\textit{E}_{cep}(\theta^m)$ feed into the downstream RS. Their model performances are inferior to UserIP-Tuning. Because the quantization module clusters more accurately and learns the causality of profile inferences.

\begin{table}[!htbp]
\setlength{\abovecaptionskip}{0cm} %# 调整间距
 \setlength{\belowcaptionskip}{-0.1cm}
    \small
   % \vspace{-1mm}
    \caption{Ablation study of quantization module}
    %\centering
    \begin{tabular*}{1\linewidth}{@{\extracolsep{\fill}} cp{0.8cm}p{0.8cm}p{0.8cm}p{0.8cm}}
% \caption{Ablation study of the VQ module on four datasets.}
\toprule[1pt]
\multirow{2}{*}{Model component} & \multicolumn{2}{c}{Clothing} & \multicolumn{2}{c}{Movies} \\ \cline{2-5} 
                                 & AUC          & Logloss       & AUC         & Logloss      \\ \hline
w.o. quantization            & 0.6201       & 0.5577        & 0.7966      & 0.4636       \\
K-Means              &   0.6211         & 0.5318        & 0.8168      & 0.4673       \\
A.F. &   0.621          & 0.5567        & 0.8175      & 0.4562     \\
UserIP-Tuning                    & \textbf{0.6272}       & \textbf{0.5033}        & \textbf{0.8189}      & \textbf{0.4517}       \\ \midrule[1pt]
\multirow{2}{*}{Model component} & \multicolumn{2}{c}{Games}    & \multicolumn{2}{c}{Yelp}   \\ \cline{2-5} 
                                 & AUC          & Logloss       & AUC         & Logloss      \\ \hline
w.o. quantization            & 0.7340       & 0.8429        & 0.7277      & 0.5511       \\
K-Means                      & 0.7297       & 0.8032      & 0.7307    & 0.5560  \\
A.F.  &   0.7301          & 0.8006        & 0.7308     & 0.5576      \\
UserIP-Tuning                    & \textbf{0.7398}       & \textbf{0.7225}        & \textbf{0.7320}      & \textbf{0.5502}       \\ \bottomrule[1pt]
\end{tabular*}
\label{table:Abla} % label放在end tabular之后
\vspace{-4mm}
\end{table}

\vspace{-0.1in}
\subsection{Parameter Analysis} \label{app:para}
\begin{figure}[!htbp]
        \setlength\abovecaptionskip{-0.1\baselineskip}
        \setlength\belowcaptionskip{-0.1\baselineskip}
	\centering
	%\hspace*{-1.25mm}
 \vspace{-1mm}
        {\subfigure[Clothing]{\includegraphics[width=0.49\linewidth]{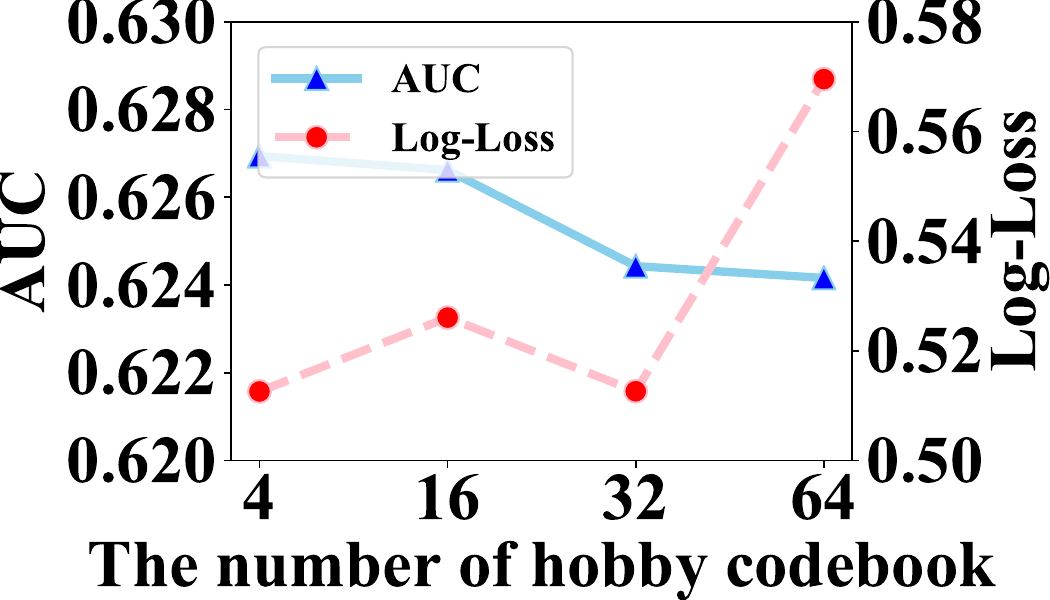}}}
        {\subfigure[Movies]{\includegraphics[width=0.49\linewidth]{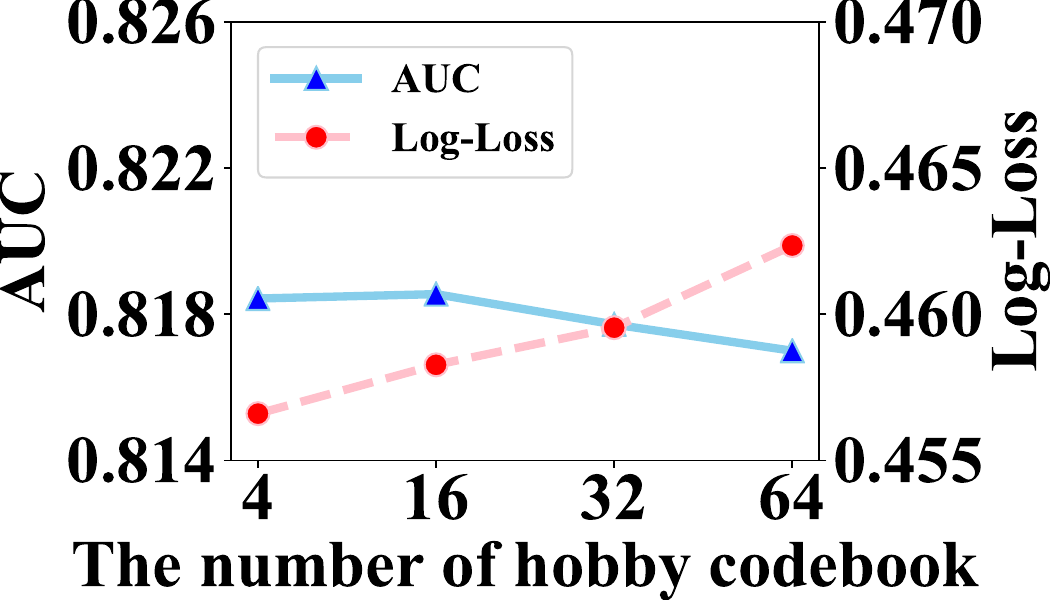}}}
        {\subfigure[Games]{\includegraphics[width=0.49\linewidth]{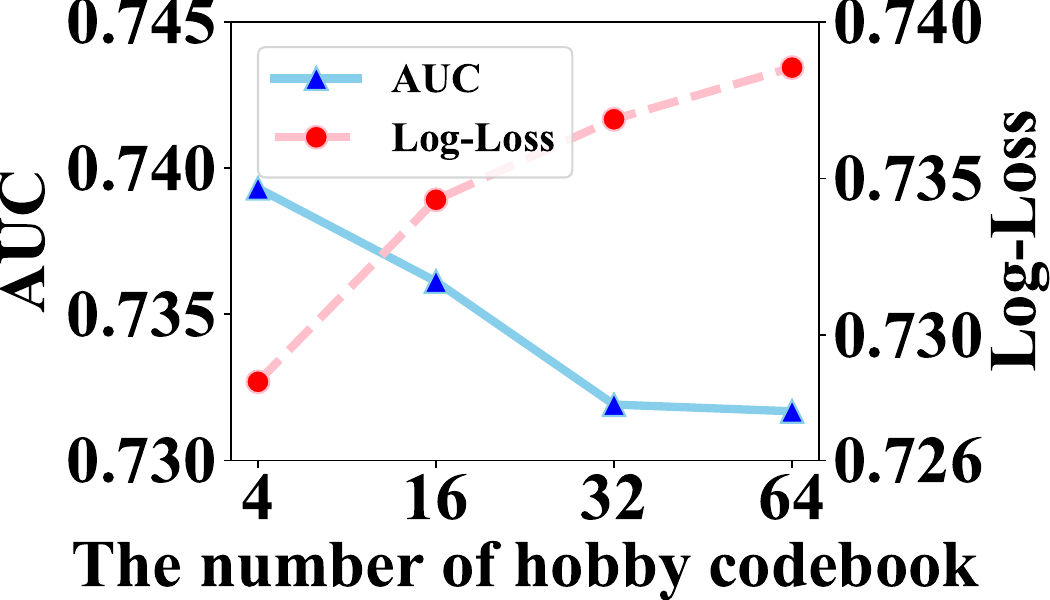}}}
        {\subfigure[Yelp]{\includegraphics[width=0.49\linewidth]{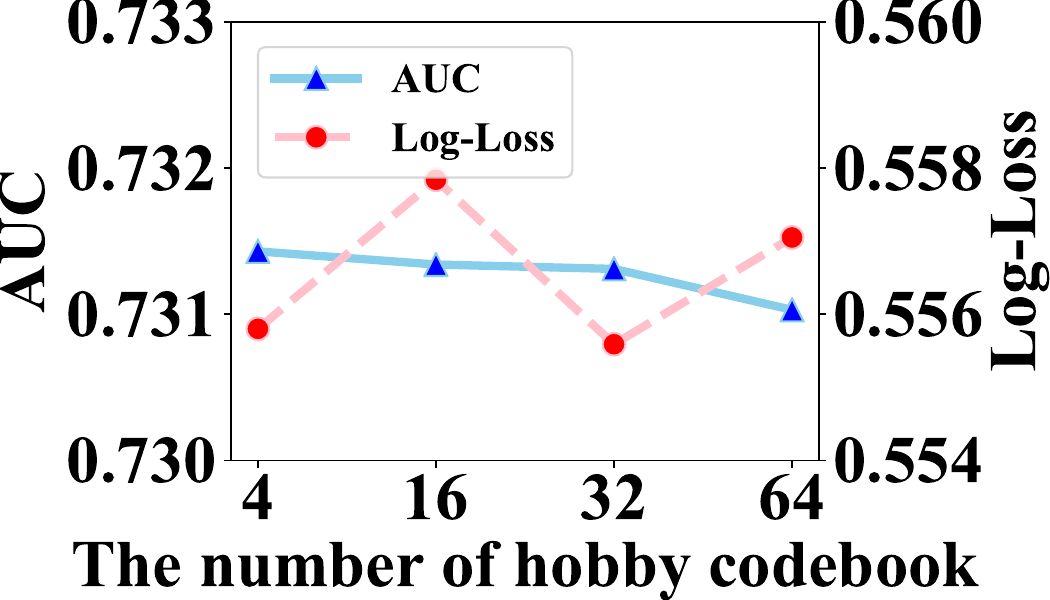}}}
	%\hspace*{7mm}
	\caption{Different quantization codebook sizes for hobby profile hyperparameter on four datasets. }
	\label{fig:k1}
		\vspace{-4mm}
\end{figure}

\begin{figure}[!htbp]
        \setlength\abovecaptionskip{-0.1\baselineskip}
        \setlength\belowcaptionskip{-0.1\baselineskip}
	\centering
 \vspace{-2mm}
	%	\hspace*{-1.25mm}
        {\subfigure{\includegraphics[width=0.49\linewidth]{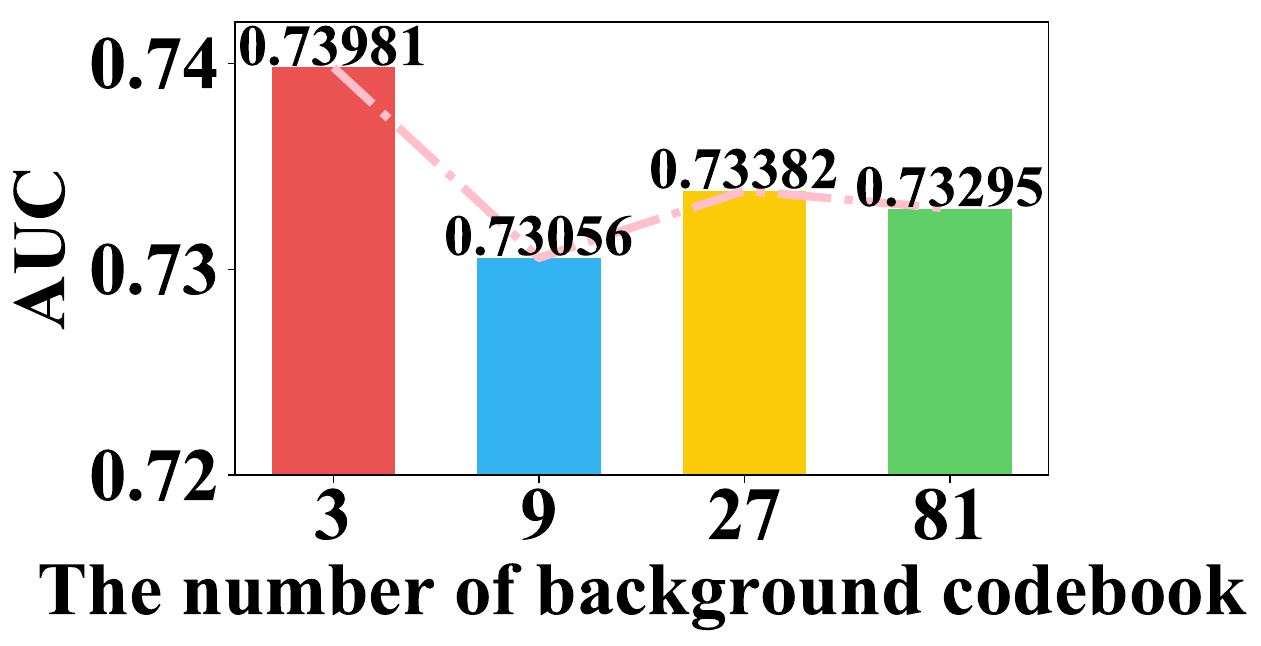}}}
        {\subfigure{\includegraphics[width=0.49\linewidth]{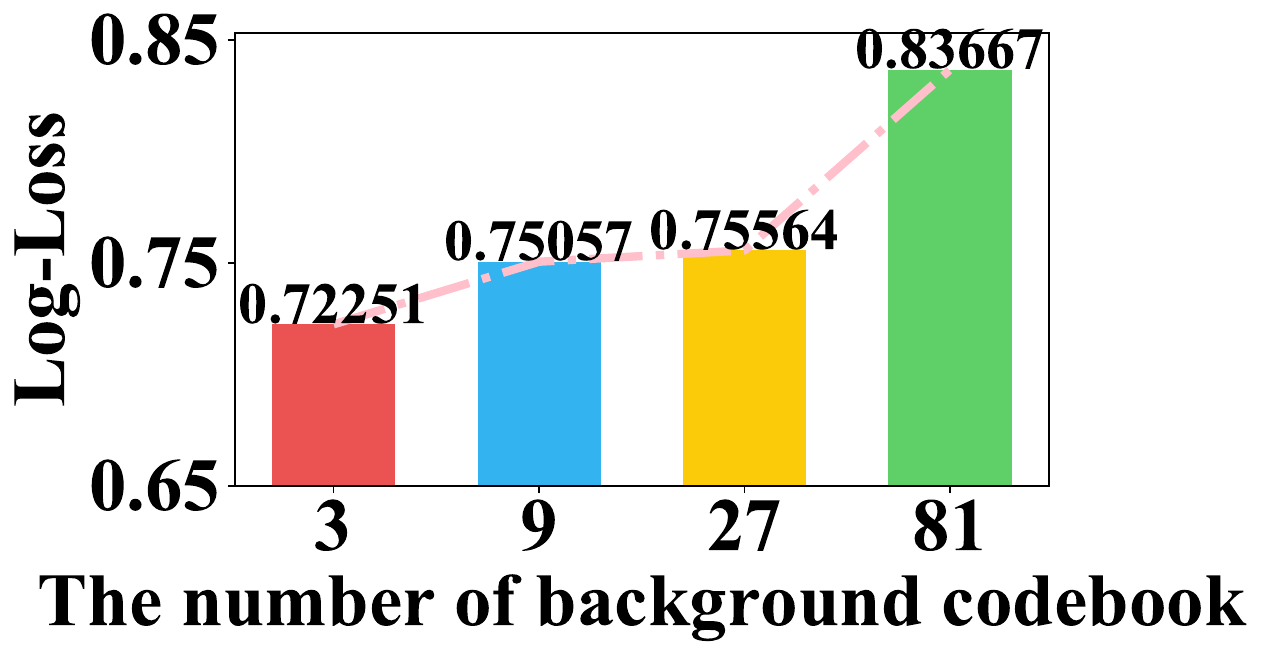}}}
	%\includegraphics[width=0.95\linewidth]{samples/Figs/hyp.pdf}
	%\hspace*{7mm}
	\caption{Background codebook sizes on Games.}
	\label{fig:hyper}
		\vspace{-2mm}
\end{figure}

We analyze the impact of the profile codebook size $K_m$ in the quantization module. 
As shown in Figure~\ref{fig:k1}, model performance declines as the hobby codebook size grows from 4 to 64: AUC decreases and Logloss increases. Excessively large codebooks produce redundant clusters, scattering user profile indices and reducing representational accuracy. We further vary the background codebook size $K_2$ from 3 to 81 (Figure~\ref{fig:hyper}). The best performance appears at size $K_2=3$, where AUC reaches 0.73931 and Logloss drops to 0.72851. Overall, smaller codebooks enable more effective latent profile clusters and yield higher recommendation accuracy.  

\vspace{-0.1in}
\subsection{Industrial Application Study} \label{sec:ind}

We evaluate UserIP-Tuning in a large-scale online advertising scenario, with results summarized in Table~\ref{tab:eff}. User profiles are inferred and stored as latent IDs. When added as features to DCN, inference speed reaches 1.21s/10,000 instances, only 0.28s slower than DCN alone. In contrast, KAR requires 4.78s, nearly four times slower than our method. Moreover, UserIP-Tuning + DCN attains an AUC of 0.7997. Our method improves efficiency and reduces memory cost for industrial recommendation tasks. 
\begin{table}[!htbp]
     \setlength{\abovecaptionskip}{0cm} %# 调整间距
     \setlength{\belowcaptionskip}{-0cm}
    \small
    %\vspace{-1mm}
    \caption{The inference time on our industrial platform}
    \centering
    \begin{tabular*}{1\linewidth}{@{\extracolsep{\fill}}ccc}
        \toprule[1pt]
        Model & Inference time (s) & AUC  \\ \midrule
        DCN & 0.93 & 0.76394\\
        KAR + DCN & 4.78  & 0.78663 \\
        UserIP-Tuning + DCN & 1.21 & 0.79972 \\\bottomrule[1pt]%\midrule
        %Rela. Improv. & 30.1075\% \\\bottomrule[1pt]
    \end{tabular*}
    \label{tab:eff}
    \vspace{-5mm}
\end{table}

\vspace{-0.1in}
\subsection{Online A/B Test}
To assess the online performance of UserIP-Tuning, we conduct a seven-day A/B test (May 13–19, 2025) on the Explore page of Huawei AppGallery. The test involves hundreds of millions of daily users, ensuring strong statistical significance. AUC is adopted as the primary evaluation metric.  

We compare our method (treatment) with a strong production baseline built on the DCN framework \cite{wang2017dcn} (control). As shown in Table~\ref{tab:abtest}, UserIP-Tuning delivers clear commercial benefits, achieving a 7.4724\% improvement in AUC.  

\begin{table}[h]
      \setlength{\abovecaptionskip}{0cm} %# 调整间距
      \setlength{\belowcaptionskip}{-0.1cm}
    \small
    \vspace{-2mm}
    \caption{Results of the online A/B experiment}
    \centering
    \begin{tabular*}{1\linewidth}{@{\extracolsep{\fill}}cc}
        \toprule[1pt]
        Model  & AUC  \\ \midrule
        Baseline  & 5.2984\%\\
        UserIP-Tuning  & 12.7708\% \\
        \midrule
        \textit{Improvement}  & 7.4724\% \\\bottomrule[1pt]%\midrule
        %Rela. Improv. & 30.1075\% \\\bottomrule[1pt]
    \end{tabular*}
    \label{tab:abtest}
    \vspace{-2mm}
\end{table}

\vspace{-0.1in}
\subsection{Case Study on Text Noise and Explainability}
\begin{figure}[!htbp]
        \vspace{-5mm}
        \setlength\abovecaptionskip{-0.15\baselineskip}
        \setlength\belowcaptionskip{-0.1\baselineskip}
        
        \small
	\centering
	%	\hspace*{-1.25mm}
        {\subfigure[Hobby profile before training]{\includegraphics[width=0.49\linewidth]{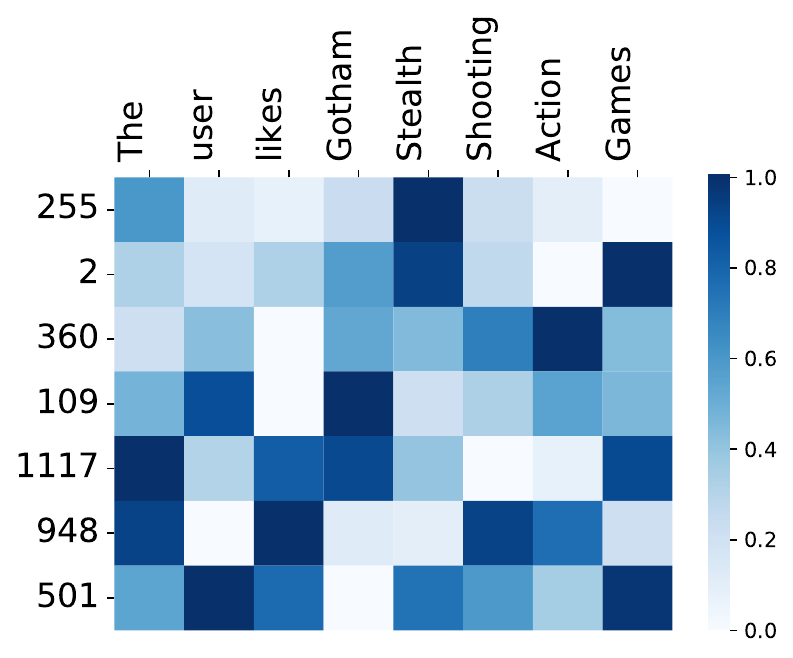}}}
        {\subfigure[Hobby profile after training]{\includegraphics[width=0.49\linewidth]{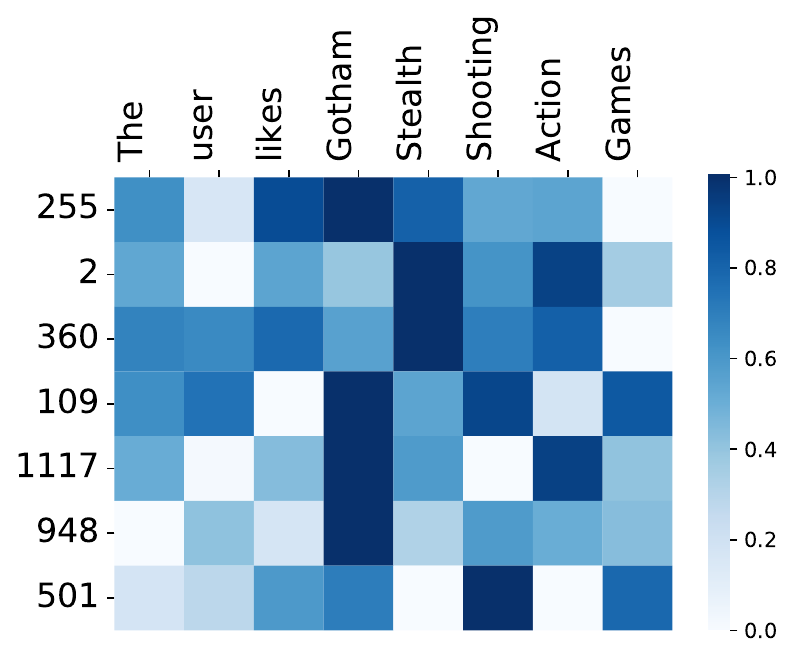}}}
        {\subfigure[Background profile before training]{\includegraphics[width=0.49\linewidth]{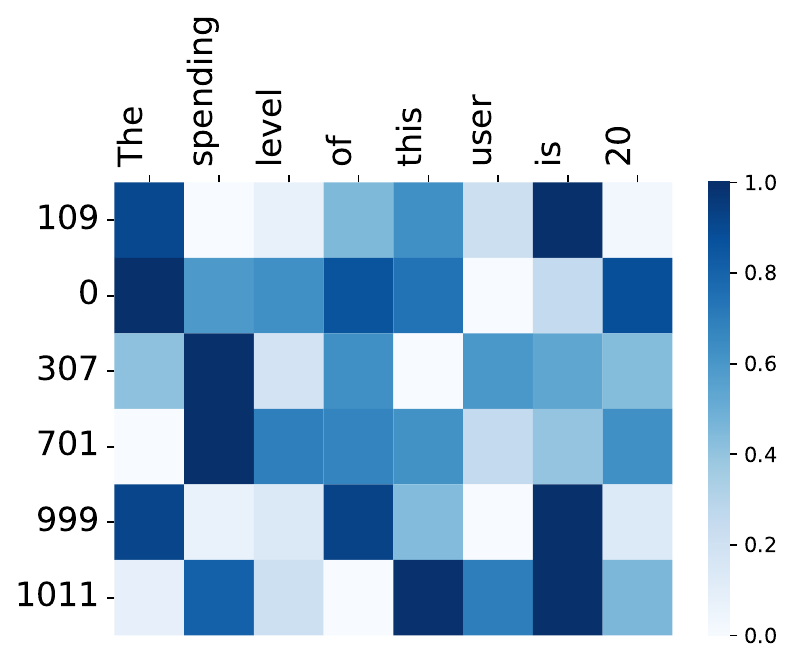}}}
        {\subfigure[Background profile after training]{\includegraphics[width=0.49\linewidth]{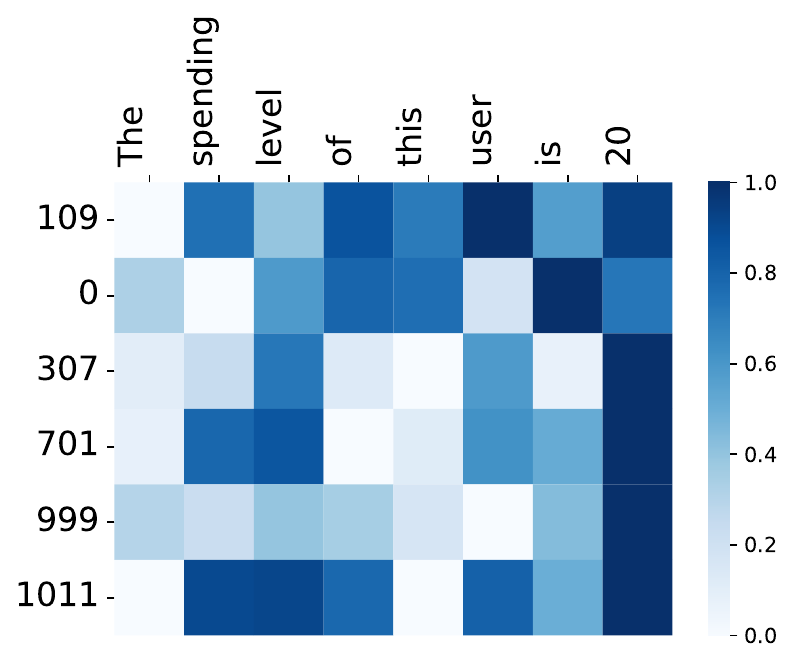}}}
	%\hspace*{7mm}
	\caption{The textual noise relationship between the users' hobby and background profiles variables and the inference text on Games dataset. }
	\label{fig:heat}
		\vspace{-4mm}
\end{figure}

We conduct a case study on textual noise and explainability by randomly selecting seven users from the Games dataset. Before training UserIP-Tuning, user hobby profile vectors are randomly initialized. We compute attention weights between these vectors and each token in the sentence ``The user likes Gotham stealth shooting action games'', and normalize them to $[0,1]$. As shown in Figure~\ref{fig:heat}(a), the textual noise ``The user'' receives higher weights. In contrast, the actual hobby-related tokens (``Gotham'', ``Stealth'', ``Shooting'', ``Action'') show weak associations due to random initialization. After training, Figure~\ref{fig:heat}(b) illustrates more substantial alignment with the hobby terms (dark blue) and reduced noise weights (light blue).  

We also analyze textual noise in background profiles by selecting six users from the Amazon Video Games dataset. Before training, attention in Figure~\ref{fig:heat}(c) mainly concentrates on noise phrases such as ``The spending level of this user is''. After training, Figure~\ref{fig:heat}(d) shows that noise weights are suppressed, while attention shifts to key profile information, indicating that textual noise is substantially mitigated. 
Finally, we verify whether latent profile variables can be effectively controlled. In Figure~\ref{fig:heat}, comparing (a) and (b), the weight on ``The user likes'' decreases markedly, while the hobby-related tokens gain much higher attention. Similarly, in (c) vs. (d), the noise phrase ``The spending level of this user is'' loses weight, whereas the critical value ``20'' receives stronger attention. 

%\subsubsection{Case Study on Textual Noise}

We conduct a case study to examine the explainability of users' latent profile variables. Specifically, we randomly select two users and retrieve cluster words with their corresponding embeddings in the LLM semantic space, as summarized in Table~\ref{tab:case}. 

\begin{table}[!htbp]
\vspace{-2mm}
\setlength\abovecaptionskip{-0\baselineskip}
        \setlength\belowcaptionskip{-0.1\baselineskip}
\small
    \caption{Hobby, background profiles retrieval}
    \centering
\begin{tabular*}{1\linewidth}{@{\extracolsep{\fill}}cc}
\toprule[1pt]
User-id & Hobby latent variable retrieval        \\ \hline
255     & {[}Gotham; Batman; Spectrum; Viper{]}  \\
136     & {[}Mortal, Magic, Gorge, expansions{]} \\ \midrule[1pt]
User-id & Background latent variable retrieval   \\ \hline
255     & {[}XBOX; LOL; WR; shoot{]}             \\
136     & {[}Genesis, Homer, Zombie, AnyWay{]}   \\ \bottomrule[1pt]
\end{tabular*}
\label{tab:case}
\vspace{-4mm}
\end{table}

For user ``255'', the retrieved words indicate a preference for superhero games such as ``Batman'', ``Spectrum'' heroes, and ``Gotham'' city, while the background terms relate to mortal and magic settings. User ``136'' shows interest in LOL: WR (League of Legends: Wild Rift) and shooting games, with background terms pointing to console use (Genesis system) and themes like Zombie or AnyWay adventures. The results show the inferred profiles are interpretable and practically useful, while effectively alleviating textual noise.  
%\vspace{-1mm}

%\vspace{-0.02in}
\section{Related Works}
\label{sec:related_work}
This section briefly introduces the previous studies about LLM-based user profile inference for RSs. 
%In recent years, the LLMs achieves remarkable in-co/ntext learning capability. 
Existing works focus on using LLMs to implicitly infer the task-related user profile information~\cite{zhang2025llmtreerec,zhang2025notellm,liu2024llm,wang2024llm4msr,zhao2023kuaisim,liu2023exploration}. 
In the field of in-context learning, \citet{xie2022explanation} design a latent document-level user profile variable through pre-trained LLMs. The LLMs implicitly perform inference while pre-trained and task-specific data distribution is hidden Markov models~\cite{luo2023recranker,li2023e4srec,hou2023large}.
\citet{min-etal-2022-metaicl} devise a new few-shot learning method where LLMs are meta-trained and conditional on training examples to recover related tasks and infer predictions.
\citet{wang2023large} construct the Bayes optimal classifier for estimating latent user profiles and selecting optimal demonstrations from training data to infer task-related topics.
\citet{chan2022data} investigate a skewed Zipfian data distribution with high burstiness in LLMs to infer the emergent in-context learning behavior.
\citet{li2023transformers} prove LLMs can apply near-optimal algorithms on traditional linear regression tasks with i.i.d. dynamic data.
To the best of our knowledge, we are the first to investigate the LLMs inference for deep RSs. \citet{wang2025crossdistillation} capture users’ preferences, contexts, and behavioral signals to enable efficient personalization in resource-constrained edge-side recommendation scenarios. 
\citet{zhang2025llm} design LLM-powered user simulators to explicitly model user preferences and sentiments, thereby generating high-fidelity behavioral data to enhance the RSs' training and evaluation.

%\vspace{-0.1in}
\section{Conclusion}
\label{sec:conclusion}
%\vspace{-5pt}
A novel LLM-based latent profile inference model, UserIP-Tuning, augments recommendation performance and training efficiency by finding user latent profile variables and lightweight profile indices. 
The causal relationship between latent profile and interaction history is considered, and a novel UserIP quantization module is designed to categorize user latent profile embedding and output users’ nearest profile indices, which extract useful collaborative signals from these embeddings. 
Our method addresses the LLM-based recommendation collaborative problem and eliminates the textual noises and out-of-range results, significantly improving training and reasoning capabilities. 
We empirically demonstrate the effectiveness, efficiency, generalizability, and explainability of UserIP-Tuning using public and industrial datasets and online A/B testing.
In the future, we will explore more industrial applications.

\begin{acks}
This research was partially supported by Hong Kong Research Grants Council's Research Impact Fund (No.R1015-23), Collaborative Research Fund (No.C1043-24GF), General Research Fund (No.11218325), Institute of Digital Medicine of City University of Hong Kong (No.9229503), Huawei (Huawei Innovation Research Program), and National Natural Science Foundation of China (Grant Nos.62502404, 72171176, 72021002 and 72471178).
\end{acks}

\section*{GenAI Usage Disclosure}
In this paper, we declare that there is no usage of GenAI in any part of the research and writing process.

\bibliographystyle{ACM-Reference-Format}
\balance
\bibliography{sample-base}

% \appendix









\end{document}